\documentclass[referee, pdflatex, sn-vancouver-num]{sn-jnl}

\usepackage{multirow}%
\usepackage{amsmath,amssymb,amsfonts}%
\usepackage{amsthm}%
\usepackage{mathrsfs}%
\usepackage[title]{appendix}%
\usepackage{xcolor}%
\usepackage{textcomp}%
\usepackage{manyfoot}%
\usepackage{booktabs}%
\usepackage{algorithm}%
\usepackage{algorithmicx}%
\usepackage{algpseudocode}%
\usepackage{listings}%

\usepackage{adjustbox}

\usepackage{graphicx}         
\usepackage[table]{xcolor}    
\usepackage{booktabs}         
\usepackage{colortbl}         
\usepackage{hyperref}         
\usepackage{makecell}

\begin{document}

\title[Article Title]{Investigations of Heterogeneity in Diagnostic Test Accuracy Meta-Analysis: A Methodological Review} 

\author*[1]{\fnm{Lukas} \sur{Mischinger}}\email{lukas.mischinger@tum.de}
\author[1]{\fnm{Angela} \sur{Ernst}}\email{angela.ernst@tum.de}
\author[1]{\fnm{Bernhard} \sur{Haller}}\email{bernhard.haller@tum.de}
\author[2]{\fnm{Alexey} \sur{Formenko}}\email{AlexeyAleksandrovic.Fomenko@mri.tum.de}
\author[3]{\fnm{Zekeriya} \sur{Aktürk}}\email{zekeriya.aktuerk@med.uni-augsburg.de}
\author[1,2]{\fnm{Alexander} \sur{Hapfelmeier}}\email{Alexander.Hapfelmeier@mri.tum.de}

\affil[1]{\orgdiv{Institute of AI and Informatics in Medicine}, \orgname{TUM School of Medicine and Health, Technical University of Munich}, \orgaddress{\street{Ismaninger Straße 22}, \city{Munich}, \postcode{81675}, \state{Bavaria}, \country{Germany}}}

\affil[2]{\orgdiv{Institute of General Practice and Health Services Research}, \orgname{TUM School of Medicine and Health, Technical University of Munich}, \orgaddress{\street{Orleansstraße 47}, \city{Munich}, \postcode{81667}, \state{Bavaria}, \country{Germany}}}

\affil[3]{\orgdiv{Institute of General Practice}, \orgname{Medical Faculty, University of Augsburg}, \orgaddress{\street{Gutenbergstraße 7}, \city{Neusäß}, \postcode{86356}, \state{Bavaria}, \country{Germany}}}


\abstract{
\textbf{Background:} 
Subgroup analyses and meta-regression are important approaches for investigating heterogeneity in diagnostic test accuracy (DTA) meta-analyses. 
However, the extent to which methodological guidance on such investigations is followed in contemporary diagnostic reviews remains unclear.
This methodological review provides an up-to-date overview of investigations of heterogeneity (IoH) in DTA meta-analyses, examining factors associated with their conduct, their frequency and characteristics, and their alignment with current methodological recommendations.
\\
\textbf{Methods:}
We included DTA meta-analyses published in 2024 that reported at least one pair of summary sensitivity and specificity. 
We excluded non-DTA reviews, narrative syntheses, studies reporting only alternative measures, and overviews of systematic reviews. 
MEDLINE (via Ovid) was searched for English-language publications, with the last search conducted in January 2025.
\\
\textbf{Results:}
From \(403\) records identified, the latest \(100\) DTA meta-analyses were included, each contributing one index test to the synthesis. Among these, IoH were reported for \(61\). 
The number of primary studies was positively associated with reporting at least one investigation (OR \(1.66\); \(p = 0.008\)), and each subgroup was substantiated by data from a median of \(6\) primary studies.
Subgroup analyses were used in \(35/61\) (\(57\%\)), while \(26/61\) (\(43\%\)) applied meta-regression, alone or in combination with subgroup analyses. Subgroup analyses involved fewer subgroup-defining variables than meta-regression (\(p < 0.001\)).
Among \(44/61\) (\(72\%\)) analyses with sufficient methodological detail to identify a statistical model used for IoH, the bivariate model was used in \(28/44\) (\(64\%\)), univariate random-effects models in \(14/44\) (\(32\%\)), and the HSROC model in \(5/44\) (\(11\%\)).
Formal tests for subgroup differences were reported in \(37/61\) (\(61\%\)); those with at least one statistically significant result investigated more subgroup variables (\(p = 0.002\)).
A study protocol was available for \(43/61\) (\(70\%\)) analyses; \(19/43\) (\(44\%\)) fully prespecified IoH. Prespecified investigations assessed fewer subgroup variables than post hoc analyses (\(p < 0.001\)).
\\
\textbf{Discussion:}
IoH were common in DTA meta-analyses and more likely when more primary studies were available. However, individual subgroups were typically based on only modest data support.
Clear reporting of statistical models and awareness of appropriate model choice for IoH were lacking in a substantial number of reviews.
To avoid undue emphasis on spurious results, greater adherence to prespecification of IoH in protocols is warranted. 
Taken together, future research should work toward closer alignment with recommendations to strengthen the reliability and transparency of IoH in diagnostic research.}


\maketitle

\section{Background}
\label{Background}
\subsection{Diagnostic test accuracy meta-analyses}
Diagnostic test accuracy (DTA) studies assess the ability of an index test to distinguish the presence or absence of a target disorder as defined by a reference standard. \cite{Bossuyt2023}
Because individual studies are limited in sample size, meta-analysis using hierarchical models is commonly applied to derive summary estimates of diagnostic performance. \cite{Leeflang2008, Knottnerus2009} 
Such analyses typically focus on sensitivity and specificity, the most frequently reported diagnostic metrics. \cite{Dahabreh2012, Honest2002} 
Depending on whether the objective is to derive a summary point or a summary line, the bivariate random-effects model \cite{Reitsma2005, Chu2006} or the hierarchical summary receiver operating characteristic (HSROC) model \cite{Rutter2001, Harbord2007} are generally considered the most appropriate approaches for meta-analysis of diagnostic test accuracy. \cite{Harbord2008, Macaskill2023, Trikalinos2012, Veroniki2022} 

\subsection{Investigation of heterogeneity}
\label{Investigation of heterogeneity}
Considerable variation in results across DTA studies is commonly observed, often exceeding what would be expected by random error alone. \cite{Lijmer2002, Macaskill2023, Deville2002}
Such between-study variability may partly be explained by differences in patient populations, study designs, how the index test was applied, or the threshold used to define a positive test result on a continuous or ordinal scale, among other factors. \cite{Lijmer2002, Irwig2002, Lijmer1999, Whiting2004, Leeflang2008}
When conducting a meta-analysis, it is advisable to explore potential sources of heterogeneity. This can enhance the scientific relevance of the findings by explaining why study results differ, and the clinical relevance by identifying in which patient groups, settings, or circumstances the index test performs variably. \cite{Thompson1994, Lijmer2002}
\\
A simple first step involves visual inspection using graphical methods, such as forest plots or summary receiver operating characteristic (sROC) curves, where symbols or colors distinguish study subsets. \cite{Macaskill2023, Lewis2001, Patel2021}
While informative for identifying patterns, such tools cannot reliably disentangle genuine heterogeneity from random variation. \cite{Cuzick2005, Buntinx2009} 
\\
A more formal approach is to perform investigations of heterogeneity (IoH), which aim to evaluate whether study-level characteristics account for some of the observed variability. Two principal approaches can be distinguished:  
(i) In subgroup analysis, study-level characteristics are categorized to divide the available studies into subgroups, for which separate summary estimates of diagnostic accuracy are then calculated. \cite{Buntinx2009, Langan2022, Deeks2024}
(ii) Meta-regression extends this concept by incorporating study-level covariates directly into a regression framework, which is statistically more efficient as it jointly analyzes all studies, accommodates continuous covariates, and permits to evaluate whether covariates explain heterogeneity. \cite{Macaskill2023, Trikalinos2012, Fu2011}
\\
Several methodological considerations have been highlighted in the literature when exploring heterogeneity in meta-analyses via subgroup analysis and meta-regression. 
Reliable and precise estimates require an adequate number of primary studies. 
As a rule of thumb for regression modeling, the Cochrane Handbook recommends at least ten primary studies per subgroup-defining covariate, cautioning that this may still be insufficient when primary studies are unevenly distributed across covariates. \cite{Deeks2024}
The Agency for Healthcare Research and Quality advises \(6\)–\(10\) studies for continuous study-level covariates and at least \(4\) for categorical covariates when primary studies are of moderate or large size. \cite{Fu2011}
Because the risk of false-positive findings increases with the number of tests for subgroup differences, appropriate adjustment for multiplicity has been recommended. \cite{Deeks2024, Deeks2023} 
Subgroups should be prespecified in study protocols and kept to a reasonable number to reduce risks of data dredging and spurious findings, although clearly labeled exploratory analyses may still be valuable for hypothesis generation. \cite{Dahabreh2016, Deeks2024, Deeks2023} 
Finally, relying on single subgrouping factors may overlook more complex effect modification; suspected effect modifiers should, where possible, be assessed jointly. \cite{Dahabreh2016} 

\subsection{Objectives}
High-quality methodological guidance is available for conducting IoH in DTA meta-analyses (e.g., in the Cochrane DTA Handbook \cite{Macaskill2023, Takwoingi2023}, practical-method reviews \cite{Trikalinos2012}, and reporting standards \cite{Veroniki2022}). A limited number of methodological surveys have evaluated how consistently these recommendations are implemented in practice.
In an early study by Dinnes et al. \cite{Dinnes2005} in \(2005\), \(133\) DTA meta-analyses were examined. Of these, \(102\) investigated potential sources of heterogeneity, with \(74\) (\(56\%\)) using subgroup analyses and \(45\) (\(34\%\)) employing some form of regression analysis. When evaluating the adequacy of the underlying data to support these investigations, the authors raised concerns about a possible overinvestigation of study characteristics.
A later review conducted in \(2012\) by the Agency for Healthcare Research and Quality \cite{Dahabreh2012} assessed \(760\) DTA meta-analyses. Among these, \(247\) (\(33\%\)) explored heterogeneity exclusively through subgroup analyses, while \(180\) (\(24\%\)) used meta-regression with or without subgroup analyses. Although an increasing tendency over time to investigate heterogeneity using subgroup analyses was observed, meta-regression methods remained relatively infrequently applied.
More recently, in \(2024\), White et al. \cite{White2024} reviewed \(242\) medical imaging DTA meta-analyses. Of these, \(122\) (\(50\%\)) conducted subgroup analyses and \(87\) (\(36\%\)) performed meta-regression. Among the \(160\) studies that used either approach, only \(13\) (\(8.1\%\)) had planned these analyses a priori. Furthermore, meta-analyses that did not investigate potential sources of heterogeneity generally provided no specific justification for this omission.
Taken together, these findings suggest that despite the availability of detailed methodological guidance, the adoption of approaches for investigating heterogeneity in practice remains far from universal.
\\
To examine this issue, we conducted an up-to-date methodological review of DTA meta-analyses, focusing on studies published in 2024 and indexed in MEDLINE. 
Our aim was to describe how heterogeneity is currently investigated by quantifying the frequency of investigations, characterizing their methodological features, and assessing factors associated with their use. The findings were interpreted in relation to existing methodological recommendations.

\section{Methods}
Where appropriate, this methodological review adhered to the PRISMA 2020 statement \cite{Page2021}. 
It was prospectively registered in PROSPERO with registration number CRD42025628658.
\subsection{Eligibility criteria}
We included original meta-analyses of diagnostic test accuracy that assessed one or more index tests against one or more reference standards and reported at least one pair of summary estimates of sensitivity and specificity. Eligible reviews had to be based on studies in humans and published in 2024 in English. To ensure a broad perspective, no restrictions were applied regarding target condition (e.g., disease) or healthcare domain.
\\
We excluded reviews that did not evaluate diagnostic test accuracy, relied solely on narrative synthesis without quantitative aggregation, reported only ranges of estimates, or presented exclusively alternative metrics such as summary hazard ratios or diagnostic odds ratios. Overviews of systematic reviews were also excluded. 

\subsection{Information sources and search strategy}
We searched MEDLINE via Ovid \cite{ovid}, using its advanced search interface. 
The strategy combined general terms related to systematic reviews of diagnostic test accuracy studies (e.g., \textit{systematic}, \textit{predict*}, \textit{diagn*}, \textit{progn*}, \textit{screen*}, \textit{test}, \textit{testing}, \textit{score}), with specific phrases such as \textit{diagnostic test accuracy}, \textit{DTA}, \textit{accuracy}, and \textit{meta-analysis}. To ensure the retrieval of studies reporting summary sensitivity and specificity, additional terms were included (\textit{sensitiv*}, \textit{specific*}, \textit{true positive}, \textit{true negative}).
\\
The search was restricted to English-language publications from 2024. The explicit implementation code, including all applied filters and limits, is available in the supplementary information. The last search was conducted in January 2025.

\subsection{Study selection and data extraction}
All references retrieved from the literature search were ordered by date of entry (newest first). 
Then the 150 most recent records were selected and imported into the PICO Portal \cite{pico}, a web-based platform for systematic review management. To assess eligibility according to the inclusion criteria, all titles and abstracts were initially screened by LM, while AH, AE, and CF each independently re-screened approximately one-third of the records. Any discrepancies in study selection were resolved through discussion until consensus was reached.
\\
For the full-text screening phase, LM independently reviewed all full-text articles to confirm eligibility. AH and AE each independently screened and reassessed half of the full texts. Disagreements were resolved through discussion to achieve consensus.
\\
Data extraction was conducted using a structured template developed in Microsoft Excel.
LM extracted data from all included full-text articles. 
To verify data extraction, AH and AE independently re-extracted data, each from half of the studies, in tandem with LM’s extraction. 
Discrepancies were resolved through discussion until consensus was reached.

\subsection{Data items}
For each DTA meta-analysis we recorded the condition studied and the total number of distinct index tests associated with summary estimates of both sensitivity and specificity. Given the strong variations in numbers of index tests due to the substantial diversity in review aims, we tried to enhance comparability by systematically selecting only one index test per review, which then served as the basis for all subsequent assessments. Inspired by the structured approach of Sun et al. \cite{Sun2011}, we prioritized the index test explicitly designated by the study authors as the principal analysis or objective. If multiple candidate tests were designated as such, we selected the one for which a summary sensitivity and specificity estimate was first reported in the abstract or, if absent there, in the results section. If no index test was designated as principal, we applied the same rule based on order of presentation. If an index test was evaluated multiple times using different statistical models (e.g., bivariate and univariate random-effects model), thresholds, or patient cohorts (e.g., training vs. validation sets), we still considered it as a single index test, provided it referred to the same underlying diagnostic approach.
\\
When an index test was identified, we collected the summary estimates of sensitivity and specificity of the meta-analysis, from which we calculated the Youden index. If reported, we captured the corresponding AUC value. 
We also recorded the number of primary studies and the total number of patients included for that specific index test meta-analysis. 
We noted whether the index test was related to artificial intelligence (AI), for instance assessing the diagnostic accuracy of deep learning algorithms. 
Additionally, we examined whether a uniform threshold for test positivity was used for meta-analysis across the included primary studies. This was determined based on explicit statements or consistent reporting of cut-off values. In cases where this information was unclear, particularly for AI-based tests with opaque classification mechanisms, the meta-analysis was classified as not using a uniform threshold. If a multiple-threshold model was employed that explicitly incorporated multiple reported thresholds from individual primary studies (see Zapf et al. \cite{Zapf2021} for more information), we considered the threshold as uniform. 
We also collected information about the study designs of the primary studies contributing to the meta-analysis of the selected index test. Specifically, we recorded whether any of the included primary studies were conducted prospectively or retrospectively, and followed a cross-sectional, cohort, case-control, or randomized controlled trial (RCT) design; if so, the respective design was marked as present for the meta-analysis.
\\
Our primary interest was whether investigations of heterogeneity had been conducted for the selected index test meta-analysis. We distinguished between two approaches:
(i) Subgroup analysis, where separate summary estimates of diagnostic accuracy are calculated for each subgroup formed by dividing the primary studies according to subgroup-defining characteristics.
(ii) Meta-regression, where these characteristics are incorporated directly as study-level covariates within a regression framework using all available primary studies.
If an IoH was reported, we counted the number of subgroup-defining variables. Furthermore, the total number of subgroups was determined by counting the subgroup-specific pairs of summary estimates of sensitivity and specificity.
We also documented whether formal subgroup comparisons yielded statistically significant results and how many, defined as reported p-values \(<0.05\) for either summary sensitivity or specificity. 
We recorded the statistical model used for deriving subgroup estimates, distinguishing between the bivariate random-effects model, the HSROC model, and two separate univariate random-effects models. When study authors only reported the software package used without specifying the underlying model, the information was classified as unclear.
Whether an IoH was pre-specified or conducted post hoc was determined from study protocols or explicit author statements. If neither source provided this information, the IoH was considered post hoc.
Finally, we evaluated whether IoH were graphically presented, noting the use of forest plots or sROC curves.

\subsection{Synthesis methods}
All analyses were performed in \textsf{R} within RStudio. \cite{R, RStudio} 
The distribution of continuous variables is presented by medians and interquartile ranges (IQRs). For categorical variables counts and percentages are provided. 
Summary tables were generated with the \textit{arsenal} package. \cite{R:arsenal} 
\\
To examine factors associated with reporting at least one IoH, we fitted univariable logistic regression models and reported odds ratios (OR) with \(95\%\) confidence intervals (CI). Continuous predictors were scaled for increased interpretability: Youden index and AUC per \(0.1\) units, number of primary studies per \(5\) studies, and number of patients on a \(log_{10}\) scale (i.e. per \(10\)-fold increase). Wald \(z\)-tests (two-sided, \(\alpha=0.05\)) were used for inference.
As sensitivity analysis, we repeated the univariable models restricted to the \(65\) meta-analyses reporting a single index test. 
To control for confounding effects, we fitted a multivariable logistic regression model with three covariates on the reporting of at least one IoH: 
To assess the relation to diagnostic performance, the Youden index was included as covariate. The AUC value was not included as it was strongly correlated with the Youden index (\(r = 0.812; \, p < 0.001\)) and not consistently reported. 
To capture the association between data support and the conduct of IoH, the number of primary studies was included, as methodological recommendations highlight its importance for obtaining robust subgroup estimates.\cite{Deeks2023, Fu2011} 
We further adjusted for case-control study design, as this design carries a risk of spectrum bias and is explicitly queried in established quality appraisal tools such as QUADAS-2.\cite{Rutjes2005, Whiting2011, Reitsma2023}
The selection of covariates was guided by methodological considerations and data availability; however, other study-level characteristics may also be relevant and could be considered in alternative model specifications.
\\
Among meta-analyses with IoH, we summarized (i) the number of subgroup-defining variables and (ii) the number of included primary studies disaggregated by subgroup characteristics, reporting mean, quartiles, minimum, maximum, and median values. Differences across categories were assessed using Mann–Whitney U or Kruskal–Wallis tests (two-sided, \(\alpha=0.05\)). 
To assess the underlying data support of subgroups, we first calculated, for each meta-analysis reporting IoH, the ratio of subgroups to subgroup-defining variables. We then divided the number of primary studies in each meta-analysis by this ratio and reported the resulting median value.
\\
Given the exploratory nature of this methodological survey, we did not adjust p-values for multiple testing, in line with our aim to describe current practices rather than formally testing hypotheses.

\section{Results}
\subsection{Study selection}
\begin{figure}[h]
    \centering
    \includegraphics[width=\textwidth]{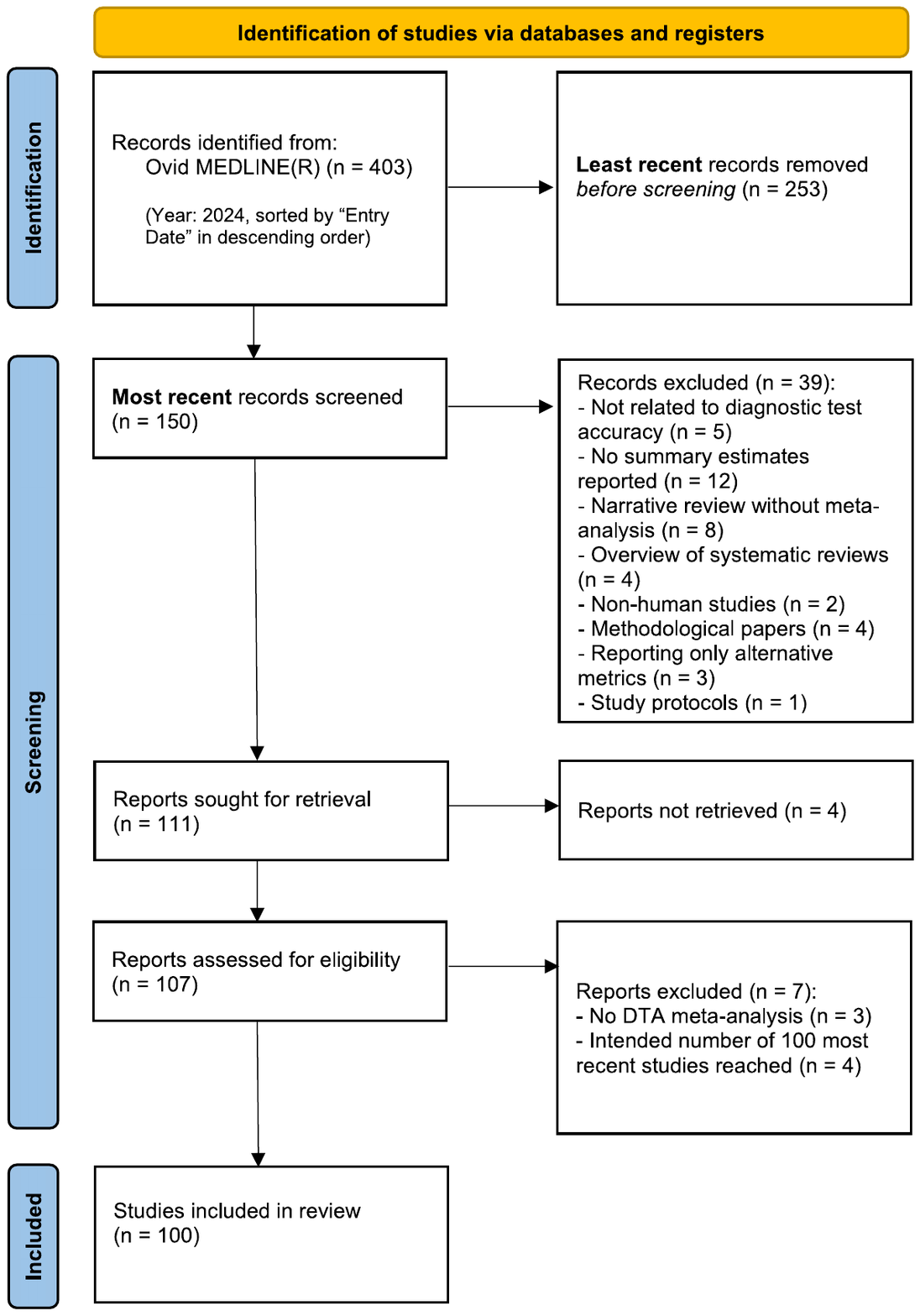}
    \caption{PRISMA flow diagram depicting the selection process of diagnostic test accuracy meta-analyses included in this review.}
    \label{Flow chart}
\end{figure}

As illustrated in Figure \ref{Flow chart}, the systematic literature search identified 403 English-language records indexed in MEDLINE (via Ovid) for the year 2024. Records were sorted by entry date in descending order, and the 150 most recent entries were selected for initial screening. Following title and abstract screening, 39 records were excluded for not reporting a DTA meta-analysis, resulting in 111 studies remaining for full-text assessment.
\\
Of these, four full-text articles could not be accessed \cite{Peng2024, Lund2024, Sikkenk2024, Woo2024}. Of the studies that were accessible, one study reported the development of a diagnostic model without conducting a DTA meta-analysis \cite{Wang2024}, one was excluded due to inclusion of only a single primary study \cite{Shabaninejad2024}, and another used a non-meta-analytic method to aggregate diagnostic accuracy results \cite{Rey2024}. 
\\
Upon reaching the predefined sample size of 100 included meta-studies, the four least recent of the remaining eligible records were not assessed further. A total of 100 original DTA meta-analyses assessing the performance of 100 distinct index tests were therefore included in the final quantitative synthesis.

\subsection{General characteristics of DTA meta-analyses}  
Table \ref{Tab1} summarizes the characteristics of the meta-analyses of the \(100\) included index tests. 
Information on AUC values, patient numbers, whether uniform cut-offs were used, and study design features of the included primary studies (prospective/retrospective, cohort, case–control, cross-sectional, randomized controlled trials) was sometimes missing or insufficiently detailed, meaning that fewer than the full set of \(100\) meta-analyses could be considered for these characteristics. 
Overall, a median of \(10\) primary studies (IQR \(7\)–\(14\); \(N = 100\)) and \(2003\) patients (IQR \(849\)–\(6255\); \(N = 79\)) was observed.  
\(20/83\) (\(24\%\)) of meta-analyses evaluated all primary studies at a uniform cut-off for test positivity.  
\(24/100\) (\(24\%\)) investigated AI-related applications, such as the DTA of deep-learning algorithms for radiomic analyses. 
The median diagnostic accuracy, as measured by the Youden index, was \(0.61\) (IQR \(0.52\)–\(0.77\); \(N = 100\)), while the median AUC was \(0.86\) (IQR \(0.81\)–\(0.91\); \(N = 60\)). 
Among the \(60\) meta-analyses that reported both diagnostic metrics, the median Youden index was \(0.59\) (IQR \(0.51\)–\(0.71\)), and the Pearson correlation between Youden index and AUC was \(r = 0.81\).
Where study design information was available, \(50/69\) (\(72\%\)) of meta-analyses included at least one prospective and \(66/72\) (\(92\%\)) at least one retrospective primary study.
Cohort (\(33/45\); \(73\%\)) and cross-sectional (\(24/49\); \(49\%\)) designs were common, whereas case-control (18/54; \(33\%\)) and randomized trials (\(1/48\); \(2\%\)) were infrequent.   

\subsection{Factors associated with investigations of heterogeneity}
Among the evaluated \(100\) meta-analyses, \(61\) (\(61\%\)) reported at least one IoH (Table \ref{Tab1}).
The amount of data available for meta-analysis was associated with the likelihood of reporting an IoH. 
In univariable logistic regression, each additional \(5\) primary studies was significantly related to a \(1.66\)-fold increase in the odds of reporting at least one IoH (OR \(1.66\), 95\% CI \(1.21\)--\(2.55\); \(p = 0.008\); \(N = 100\)). 
Meta-analyses that reported at least one IoH had a higher median number of primary studies (\(12\), IQR \(10\)--\(15\)) compared with those that did not (\(6\), IQR \(4\)--\(11\)). 
A similar but more variable trend was observed for the number of patients: a \(10\)-fold increase corresponded to \(1.69\) times higher odds (OR \(1.69\), 95\% CI \(0.86\)--\(3.68\); \(p = 0.148\); \(N = 79\)), with medians of \(2759\) (IQR \(1200\)--\(6708\)) versus \(1250\) (IQR \(674\)--\(5783\)) patients, respectively (Table \ref{Tab1}).  
\\
Across study designs, only the inclusion of at least one case-control study showed a significant association with reporting an IoH. 
The presence of any case-control study resulted in \(3.91\) times higher odds of reporting an IoH (95\% CI \(1.15\)--\(15.94\); \(p = 0.038\); \(N = 54\)), with such studies present in \(14/31\) (\(45\%\)) of meta-analyses with IoH compared with \(4/23\) (\(17\%\)) of those without. 
Other designs, including prospective, retrospective, cohort, cross-sectional, and randomized controlled trials, showed no statistically significant relationship (Table \ref{Tab1}).
\\
Regarding diagnostic accuracy measures, no clear differences were observed between meta-analyses with and without IoH. 
The Youden index showed only a weak trend in the odds ratio (OR \(1.19\) per \(0.1\) increase, 95\% CI \(0.99\)--\(1.46\); \(p = 0.073\); \(N = 100\)), with medians differing slightly (\(0.64\), IQR \(0.54\)--\(0.79\) vs. \(0.59\), IQR \(0.49\)--\(0.72\)). The AUC values also displayed a positive trend in the odds ratio (OR \(1.39\) per \(0.1\) increase, 95\% CI \(0.82\)--\(2.48\); \(p = 0.229\); \(N = 60\)), whereas medians showed little differences (\(0.85\), IQR \(0.80\)--\(0.93\) vs. \(0.86\), IQR \(0.82\)--\(0.90\)) (Table \ref{Tab1}). 
\\
Whether the index test was AI-based showed only a weak association with the reporting of IoH (OR \(1.09\), 95\% CI \(0.43\)--\(2.88\); \(p = 0.863\); \(N = 100\)). 
The use of a uniform cut-off showed higher odds (OR \(2.11\), 95\% CI \(0.72\)--\(7.14\); \(p = 0.196\); \(N = 83\)), but the confidence interval was wide, providing no clear evidence of an association.
Meta-analyses of reviews reporting more than one distinct index test had lower odds of reporting an IoH (OR \(0.54\), 95\% CI \(0.23\)--\(1.25\); \(p = 0.152\); \(N = 100\)) (Table \ref{Tab1}). 
\\
To assess whether our systematic approach of selecting only a single index test per DTA meta-analysis could have introduced bias, we conducted a sensitivity analysis restricted to DTA meta-analyses that reported only one index test with summary diagnostic estimates (\(N = 65\)). 
All covariates retained their direction of effect. Notable changes in effect size were observed for the number of patients, cohort design, and case-control design, which could be due to an increased random error associated with the reduced sample size (Table \ref{SensAnalysis}). 
\\
We conducted a multivariable logistic regression to jointly evaluate the relation of the Youden index, the number of included primary studies, and the presence of a case-control design on the likelihood of reporting at least one IoH. In this model, the associations for the Youden index (OR \(1.11\), 95\% CI \(0.84\)--\(1.51\)) and for case-control design (OR \(3.11\), 95\% CI \(0.78\)--\(14.3\); \(N = 54\)) were broadly consistent with the univariable results, while the conditional association with the number of studies was stronger, with the odds ratio increasing to \(2.83\) (95\% CI \(1.44\)--\(6.74\); \(p = 0.008\); \(N = 54\)) (Table \ref{MVAnalysis}).

\begin{sidewaystable}[h]
\caption{Descriptive characteristics and univariable logistic regression results for diagnostic test accuracy meta-analyses with vs.\ without investigations of heterogeneity}
\label{Tab1}
\begin{tabular}{@{}lcccrrrr@{}}
\toprule
 & IoH Present (N=61) & IoH Absent (N=39) & Total (N=100) & OR & 95\% CI & p-value$^{*}$ & N\\
\midrule
\textbf{Youden index} &  &  &  & 1.19$^{a}$ & (0.99–1.46) & 0.073 & 100\\
Median & 0.64 & 0.59 & 0.61 &  &  &  & \\
Q1, Q3 & (0.54, 0.79) & (0.49, 0.72) & (0.52, 0.77) &  &  &  & \\
\midrule
\textbf{AUC value} &  &  &  & 1.39$^{a}$ & (0.82–2.48) & 0.229 & 60\\
Median & 0.85 & 0.86 & 0.86 &  &  &  & \\
Q1, Q3 & (0.8, 0.93) & (0.82, 0.9) & (0.81, 0.91) &  &  &  & \\
\midrule
\textbf{Number primary studies} &  &  &  & 1.66$^{b}$ & (1.21–2.55) & \textbf{0.008} & 100\\
Median & 12 & 6 & 10 &  &  &  & \\
Q1, Q3 & (10, 15) & (4, 11) & (7, 14) &  &  &  & \\
\midrule
\textbf{Number patients} &  &  &  & 1.69$^{c}$ & (0.86–3.68) & 0.148 & 79\\
Median & 2759 & 1250 & 2003 &  &  &  & \\
Q1, Q3 & (1200, 6708) & (674, 5783) & (849, 6255) &  &  &  & \\
\midrule
\textbf{$>$1 Index tests (vs.\ 1)} & 18/61 (30\%) & 17/39 (44\%) & 35/100 (35\%) & 0.54 & (0.23–1.25) & 0.152 & 100\\
\midrule
\textbf{AI-related study (ref.\ no)} & 15/61 (25\%) & 9/39 (23\%) & 24/100 (24\%) & 1.09 & (0.43–2.88) & 0.863 & 100\\
\textbf{Uniform cut-off (ref.\ no)} & 15/52 (29\%) & 5/31 (16\%) & 20/83 (24\%) & 2.11 & (0.72–7.14) & 0.196 & 83\\
\midrule
\textbf{Prospective study (ref.\ no)} & 32/43 (74\%) & 18/26 (69\%) & 50/69 (72\%) & 1.29 & (0.43–3.8) & 0.641 & 69\\
\textbf{Retrospective study (ref.\ no)} & 43/46 (93\%) & 23/26 (88\%) & 66/72 (92\%) & 1.87 & (0.32–10.82) & 0.465 & 72\\
\midrule
\textbf{Cohort study (ref.\ no)} & 21/26 (81\%) & 12/19 (63\%) & 33/45 (73\%) & 2.45 & (0.64–9.98) & 0.193 & 45\\
\textbf{Case-control study (ref.\ no)} & 14/31 (45\%) & 4/23 (17\%) & 18/54 (33\%) & 3.91 & (1.15–15.94) & \textbf{0.038} & 54\\
\textbf{Cross-sectional study (ref.\ no)} & 12/26 (46\%) & 12/23 (52\%) & 24/49 (49\%) & 0.79 & (0.25–2.42) & 0.674 & 49\\
\textbf{RCT study (ref.\ no)} & 1/29, (3\%) & 0/19 (0\%) & 1/48 (2\%) &  &  &  & 48\\
\bottomrule
\end{tabular}
\footnotetext{$^{*}$Wald z-test for effect estimates in univariable logistic regression model. $^{a}$Per 0.1 increase. $^{b}$Per 5 additional primary studies. 
$^{c}$Based on a log$_{10}$ transformation of the number of patients; odds ratio reflects a 10-fold increase.
Denominators reflect the number of non-missing information for each characteristic. 
Study design characteristics represent the proportion of index test meta-analyses for which at least one underlying primary study was of the respective design.
All tests were exploratory and no multiplicity adjustment was considered. IoH: investigation of heterogeneity, RCT: randomized controlled trial, AUC: area under the curve, OR: odds ratio.}
\end{sidewaystable} 

\subsection{Characteristics of reported investigations of heterogeneity}
Among the \(61\) meta-analyses that reported IoH, not all subgroup-related characteristics were consistently described. In several cases, details on testing for subgroup differences, the prespecification status of the IoH, or the statistical model used were missing or unclear. As a result, fewer than the full set of \(61\) meta-analyses contributed data for these subgroup characteristics.
The median number of subgroup-defining variables was \(4\) (IQR \(1\)--\(6\); Table \ref{Tab2}), and the median number of included primary studies was \(12\) (IQR \(10\)--\(15\); Table \ref{Tab3}). 
The median ratio of the number of subgroups investigated to the number of subgroup-defining variables was \(1{:}2\), indicating that, on the median, each subgroup-defining variable comprised two subgroups.
The median ratio of number of primary studies by this ratio of subgroups per variable was \(1{:}6\), suggesting that each subgroup was substantiated by data from six primary studies on average.
\\
Formal tests of subgroup differences in DTA were reported in \(37/61\) (\(61\%\)).
Of these, \(25/37\) (\(68\%\)) identified at least one statistically significant difference. The number of subgroup-defining variables evaluated was substantially higher in meta-analyses with significant findings compared with those without (median \(5\), IQR \(4\)--\(9\) vs. median \(1.5\), IQR \(1\)--\(4.5\); \(p = 0.002\)) (Table \ref{Tab2}), whereas the number of studies included was comparable between the two groups (median \(11\), IQR \(9\)--\(13\) vs. median \(12\), IQR \(10\)--\(14\); \(p = 0.613\)) (Table \ref{Tab3}).
\\
The types of IoH varied: \(35/61\) (\(57\%\)) applied subgroup analysis (SGA) alone, \(8/61\) (\(13\%\)) used only meta-regression (MR), and \(18/61\) (\(30\%\)) combined both approaches. The number of subgroup-defining variables differed significantly between these types (\(p < 0.001\)), with the highest counts observed for meta-regression (median \(7.5\), IQR \(5\)--\(10\)), followed by the combination of meta-regression with subgroup analysis (median \(5\), IQR \(4\)--\(7\)), while subgroup analysis alone had the lowest counts (median \(1\), IQR \(1\)--\(3.5\)) (Table \ref{Tab2}). In contrast, the number of primary studies was comparable across the three types (SGA: median \(11\), IQR \(10\)--\(15\); MR: median \(11\), IQR \(8\)--\(14\); SGA + MR: median \(12\), IQR \(10\)--\(15\); \(p = 0.808\)) (Table \ref{Tab3}).
\\
For \(43/61\) (\(70\%\)) the specification status of IoH were accessible. Of those \(19/43\) (\(44\%\)) reported only prespecified IoH, \(14/43\) (\(33\%\)) reported only post hoc investigations, and \(10/43\) (\(23\%\)) included both. The number of subgroup-defining variables differed significantly between these categories (\(p < 0.001\)), with the highest counts in cases reporting both prespecified and post hoc investigations (median \(8\), IQR \(5.5\)--\(9\)), followed by post hoc alone (median \(5\), IQR \(3\)--\(6\)), and prespecified alone (median \(1\), IQR \(1\)--\(2\)) (Table \ref{Tab2}). In contrast, the number of primary studies was comparable across the three groups (prespecified: median \(12\), IQR \(10\)--\(13\); post hoc: median \(12\), IQR \(8\)--\(21\); both: median \(13\), IQR \(10\)--\(15\); \(p = 0.845\)) (Table \ref{Tab3}).
\\
When considering the graphical presentation of IoH, \(28/61\) (\(46\%\)) index test meta-analyses presented their results at least in part graphically. Forest plots alone were used in \(15/61\) (\(25\%\)) cases, sROC curves alone in \(5/61\) (\(8\%\)) cases, and both forest plots and sROC curves in \(10/61\) (\(16\%\)) cases. 
The number of subgroup-defining variables tended to vary across presentation formats: sROC curves alone showed the highest number of variables (median \(4\), IQR \(3\)--\(5\)), followed by forest plots alone (median \(2\), IQR \(1\)--\(4\)), while combined forest plot and sROC displays were used for the lowest counts (median \(1\), IQR \(1\)--\(5.5\)) (Table \ref{Tab2}). 
In contrast, the number of primary studies varied little across formats (sROC: median \(12\), IQR \(11\)--\(24\); forest plot: median \(10\), IQR \(8\)--\(29\); forest plot + sROC: median \(13\), IQR \(10\)--\(13\); \(p = 0.747\)) (Table \ref{Tab3}).
\\
The statistical model used for IoH was identifiable in \(44/61\) (\(72\%\)) meta-analyses of index tests. As in some cases authors applied more than one model across different subgroups, this category is not mutually exclusive. 
Most obtained summary sensitivity and specificity using the bivariate model (\(28/44\), \(64\%\)) or two univariate random-effects models (\(14/44\), \(32\%\)), whereas the HSROC model was applied less frequently (\(5/44\), \(11\%\)).
Among those using the bivariate model, only \(7/28\) (\(25\%\)) applied a uniform cut-off for test positivity across primary studies. 
The number of subgroup-defining variables differed across modeling approaches: meta-analyses using univariate models investigated the highest number of variables (median \(5.5\), IQR \(2.25\)--\(9.25\)), followed by those using the bivariate model (median \(3\), IQR \(1\)--\(7\)), while analyses based on the HSROC model typically included few variables (median \(1\), IQR \(1\)--\(4\)). In contrast, the number of primary studies contributing to these analyses was largely comparable across model types (median \(13\), IQR \(9\)--\(15\) for the univariate model; median \(12\), IQR \(10\)--\(16\) for the bivariate model; and median \(10\), IQR \(10\)--\(12\) for the HSROC model) (Table \ref{Tab2}, Table \ref{Tab3}).

\begin{sidewaystable}[h]
\caption{Frequency of meta-analyses reporting at least one investigation of heterogeneity and corresponding number of subgroup-defining variables disaggregated by subgroup characteristics}
\label{Tab2}
\begin{tabular}{@{}lcccccccr@{}}
\toprule
& \multicolumn{1}{c}{\textbf{Reporting IoH}}
& \multicolumn{7}{c}{\textbf{Number of subgroup-defining variables}} \\
\cmidrule(l){2-2}\cmidrule(l){3-9}
& n/N (\%) & Mean & Min & 1st Quartile & Median & 3rd Quartile & Max & p-value$^{*}$\\
\midrule
\addlinespace
$\geq$1 Significant IoH & 25/37 (68\%) & 6.6 & 3 & 4 & 5 & 9 & 17 & \textbf{0.002}\\
No Significant IoH & 12/37 (32\%) & 2.9 & 1 & 1 & 1.5 & 4.5 & 7 & \\
\midrule
\addlinespace
SGA & 35/61 (57\%) & 2.5 & 1 & 1 & 1 & 3.5 & 10 & $\mathbf{<0.001}$\\
MR + SGA & 18/61 (30\%) & 5.9 & 2 & 4 & 5 & 7 & 17 & \\
Meta-Regression & 8/61 (13\%) & 7.2 & 3 & 5 & 7.5 & 10 & 10 & \\
\midrule
\addlinespace
Prespecified IoH & 19/43 (44\%) & 2.0 & 1 & 1 & 1 & 2 & 7 & $\mathbf{<0.001}$\\
Post hoc IoH & 14/43 (33\%) & 5.1 & 1 & 3 & 5 & 6 & 10 & \\
Prespecified + Post hoc & 10/43 (23\%) & 8.1 & 4 & 5.5 & 8 & 9 & 17 & \\
\midrule
\addlinespace
Forest Plot & 15/61 (25\%) & 2.6 & 1 & 1 & 2 & 4 & 10 & \textbf{0.028}\\
Forest + sROC & 10/61 (16\%) & 2.9 & 1 & 1 & 1 & 5.5 & 7 & \\
sROC Curve & 5/61 (8\%) & 4.0 & 3 & 3 & 4 & 5 & 5 & \\
No Forest or sROC & 31/61 (51\%) & 5.3 & 1 & 2 & 5 & 8 & 17 & \\
\midrule
\addlinespace
Bivariate Model & 28/44 (64\%) & 4.3 & 1 & 1 & 3 & 7 & 10 & NA$^{a}$\\
Univariate Model & 14/44 (32\%) & 5.9 & 1 & 2.25 & 5.5 & 9.25 & 17 & \\
HSROC Model & 5/44 (11\%) & 2.8 & 1 & 1 & 1 & 4 & 7 & \\
Bivariate (UC) & 7/28 (25\%) & 4.4 & 1 & 1.5 & 3 & 7 & 10 & \\
\midrule
\addlinespace
Total & 61/61 (100\%) & 4.1 & 1 & 1 & 4 & 6 & 17 & \\
\bottomrule
\end{tabular}
\footnotetext{Denominators reflect the number of subgroup meta-analyses with non-missing information for each characteristic. $^{*}$p-values from Mann–Whitney U or Kruskal–Wallis tests comparing the distribution of the number of subgroup-defining variables across subgroup characteristics. $^{a}$Not mutually exclusive. IoH: investigation of heterogeneity, SGA: subgroup analysis, MR: meta-regression, sROC: summary receiver operating characteristic, UC: uniform cut-off across primary studies.}
\end{sidewaystable}

\begin{sidewaystable}[h]
\caption{Number of meta-analyses reporting at least one investigation of heterogeneity and corresponding number of primary studies disaggregated by subgroup characteristics}
\label{Tab3}
\begin{tabular}[t]{@{}lcccccccr@{}}
\toprule
& \multicolumn{1}{c}{\textbf{Meta-analyses}}
& \multicolumn{7}{c}{\textbf{Number of primary studies}} \\
\cmidrule(l){2-2}\cmidrule(l){3-9}
 & n & Mean & Min & 1st Quartile & Median & 3rd Quartile & Max & p-value$^{*}$ \\
\midrule
\addlinespace
$\geq$1 Significant IoH & 25 & 16.3 & 6 & 9 & 11 & 13 & 68 & 0.613\\
No Significant IoH & 12 & 14.6 & 8 & 10 & 12 & 14 & 45 & \\
\midrule
\addlinespace
SGA & 35 & 15.0 & 5 & 10 & 11 & 15 & 56 & 0.808\\
MR + SGA & 18 & 17.6 & 6 & 10 & 12 & 15 & 45 & \\
Meta-Regression & 8 & 18.0 & 7 & 8 & 11 & 14 & 68 & \\
\midrule
\addlinespace
Prespecified IoH & 19 & 12.1 & 5 & 10 & 12 & 13 & 24 & 0.845 \\
Post hoc IoH & 14 & 17.6 & 5 & 8 & 12 & 21 & 54 & \\
Prespecified + Post hoc & 10 & 16.9 & 6 & 10 & 13 & 15 & 56 & \\
\midrule
\addlinespace
Forest Plot & 15 & 19.6 & 5 & 8 & 10 & 29 & 56 & 0.747\\
Forest + sROC & 10 & 12.4 & 8 & 10 & 13 & 13 & 20 & \\
sROC Curve & 5 & 18.0 & 8 & 11 & 12 & 24 & 35 & \\
No Forest or sROC & 31 & 15.5 & 6 & 10 & 12 & 15 & 68 & \\
\midrule
\addlinespace
Bivariate Model & 28 & 14.8 & 5 & 10 & 12 & 16 & 68 & NA$^{a}$\\
Univariate Model & 14 & 15.6 & 6 & 9 & 13 & 15 & 41 & \\
HSROC Model & 5 & 11.0 & 8 & 10 & 10 & 12 & 15 & \\
Bivariate (UC) & 7 & 11.4 & 5 & 9 & 11 & 14 & 18 & \\
\midrule
\addlinespace
Total & 61 & 16.2 & 5 & 10 & 12 & 15 & 68 & \\
\bottomrule
\end{tabular}
\footnotetext{$^{*}$p-values from Mann–Whitney U or Kruskal–Wallis tests comparing the distribution of the number of primary studies across subgroup characteristics. $^{a}$Not mutually exclusive. IoH: investigation of heterogeneity, SGA: subgroup analysis, MR: meta-regression, sROC: summary receiver operating characteristic, UC: Uniform cut-off across primary studies.}
\end{sidewaystable}

\section{Discussion}
\subsection{Key Results}
Investigations of heterogeneity using subgroup analyses or meta-regression remain common in DTA meta-analyses. 
Their use was more likely in meta-analyses including larger numbers of primary studies, suggesting that stronger data support facilitates their conduct. 
However, among meta-analyses that performed IoH, each subgroup was typically substantiated by six primary studies, indicating modest underlying data support.
\\
A considerable proportion of reviews provided insufficient information about the statistical model for IoH, often reporting only the software used. Among reviews with sufficient information, the bivariate model was most frequently applied, yet a substantial number of analyses still relied on separate univariate models despite clear methodological guidance favoring hierarchical approaches. Clearer reporting and greater methodological awareness are therefore needed.
\\
Significant DTA differences were more often observed in analyses that tested larger numbers of subgroups, raising concerns that these findings may be results of multiplicity rather than due to true underlying effect modification.
Furthermore, while most meta-analyses with IoH provided a protocol, fewer than half were fully prespecified. To reduce spurious findings, future research should prespecify IoH in protocols rather than relying on extensive exploratory comparisons. 

\subsection{Determinants of investigations of heterogeneity}
In this systematic review, \(100\) diagnostic test accuracy meta-analyses were included, each contributing a single index test to the synthesis. Investigations of heterogeneity (IoH) were reported in \(61\%\) of the meta-analyses, with \(35\%\) relying exclusively on subgroup analyses and \(26\%\) applying meta-regression, either alone or in combination with subgroup analyses.
These proportions are highly comparable to those reported by Dahabreh et al.\cite{Dahabreh2012} in \(2012\), where IoH were conducted in \(57\%\) of meta-analyses overall, including \(33\%\) using subgroup analyses and \(24\%\) applying meta-regression alone or in combination.
However, the observed frequencies are somewhat lower than those reported in the earlier review by Dinnes et al.\cite{Dinnes2005} in \(2005\), with \(56\%\) employing subgroup analyses and \(34\%\) using meta-regression. They are also lower than the proportions reported in the most recent review by White et al.\cite{White2024} in \(2024\), where \(50\%\) of meta-analyses conducted subgroup analyses and \(36\%\) applied meta-regression.
\\
The number of primary studies included in the meta-analysis showed a significant association with the reporting of IoH in univariable analysis (OR \(1.66\) per \(5\) additional studies; \(p = 0.008\)), which remained consistent in the multivariable and sensitivity analyses, supporting the robustness of this finding.
IoH generally require a sufficient number of studies to be meaningful \cite{Deeks2023, Deeks2024}, and authors appear to perform them primarily when enough studies are available. The association with patient numbers (OR \(1.69\), 95\% CI \(0.86\)–\(3.68\); \(p = 0.148\)) was less precisely estimated.
The inclusion of case-control studies also showed a significant association in univariable analysis (OR \(3.91\); \(p = 0.038\)), although this association was less consistent in the multivariable and sensitivity analysis.
Case-control designs are frequently viewed with caution in diagnostic accuracy research, as the artificial separation of ``cases'' and ``controls'' can introduce spectrum bias and lead to inflated estimates of sensitivity and specificity compared with studies conducted in representative clinical populations. \cite{Rutjes2005, Knottnerus2009, Lijmer1999}  
This concern is reflected in established quality appraisal tools such as QUADAS-2, which explicitly queries whether a case--control design was avoided when assessing the risk of bias in patient selection. \cite{Whiting2011, Reitsma2023} 
This association might indicate that authors are more likely to explore heterogeneity when potentially biased case-control designs are present, which aligns with methodological recommendations.
\\
Previous work on subgroup analyses in clinical trials suggested that such analyses were more frequently reported when the primary analysis did not reach statistical significance, raising concerns that they may have been conducted in search of positive findings. \cite{Brand2021} Drawing on this observation, we hypothesized that lower diagnostic performance (Youden index or AUC) might similarly increase the likelihood of IoH in diagnostic test accuracy meta-analyses. However, our evaluation did not provide statistical evidence to support this hypothesis.

\subsection{Data support for investigations of heterogeneity}
Particularly in the context of subgroup analyses, previous authors have noted that the frequent lack of adequate sample size requirements may lead to imprecise estimates of sensitivity and specificity, often with wide confidence intervals. \cite{Bachmann2006}
For regression modeling, the Cochrane Handbook recommends at least ten studies per subgroup-defining covariate, while cautioning that this may still be insufficient when primary studies are unevenly distributed across covariates. \cite{Deeks2024}
The Agency for Healthcare Research and Quality provides slightly more liberal guidance, suggesting at least \(6\)--\(10\) studies for continuous study-level covariates and a minimum of \(4\) for categorical covariates when primary studies are of moderate or large size. \cite{Fu2011}
In a review from 2005, Dinnes et al. \cite{Dinnes2005} calculated the ratio of median number of subgroup-defining variables to the median number of primary studies of \(1{:}6\), suggesting that on average one variable was investigated for every six studies included. Further, they note, that in their review only \(38\%\) of meta-analyses complied with the recommendation to have at least \(10\) studies for every characteristic investigated, concluding that this might be a possible indication of overinvestigation of study characteristics.
\\
In our findings the overall median number of subgroup-defining variables was \(4\) (IQR \(1\)--\(6\)) and the corresponding median number of included primary studies was \(12\) (IQR \(10\)--\(15\)).
Subgroup analyses typically involved only a few subgroup-defining variables, whereas meta-regression was associated with substantially higher numbers of variable comparisons (median \(1\) for SGA, \(5\) for MR+SGA, and \(7.5\) for MR; \(p < 0.001\)), despite negligible differences in the median number of primary studies included (median \(11\) for SGA, \(11\) for MR, and \(12\) for MR+SGA; \(p = 0.808\)).
Having assessed the number of subgroups, we calculated for each meta-analysis the ratio of primary studies to subgroups per variable.
The resulting adjusted ratios had a median of \(1{:}6\), suggesting that each meta-analysis of a subgroup was typically substantiated by data from about six primary studies. 
For subgroup analyses the adjusted median ratio was \(1{:}5\), and for meta-regressions between \(1{:}6\) and \(1{:}7\).
Taking again the Cochrane Handbook recommendation of at least ten primary studies per subgroup-defining variable as a reference, and assuming that each variable comprises two subgroups, this corresponds to approximately five primary studies per subgroup. In our review, this lower bound appeared to be met on the median, suggesting no indication of overinvestigation.

\subsection{Statistical models} 
Information on the statistical models used for IoH could be extracted for only \(44/61\) (\(72\%\)) of the meta-analyses. Most of the remaining reviews only reported the statistical software used to derive summary estimates, without detailing the methodological choices, which made it difficult to determine the statistical model given the multiple options within each package. In addition, the categories were not mutually exclusive, as some authors applied different models across subgroups within the same review.
Where sufficient information was available, the bivariate model was applied most frequently (\(28/44\); \(64\%\)), while a substantial proportion of meta-analyses still relied on separate univariate random-effects models (\(14/44\); \(32\%\)), and the HSROC model was rarely used (\(5/44\); \(11\%\)). 
Methodological guidance generally favors hierarchical approaches such as the bivariate or HSROC models over univariate models, as they jointly estimate sensitivity and specificity and account for their correlation. \cite{McGrath2016, Leeflang2008, Macaskill2023} Although univariate models have been proposed as a pragmatic alternative in the context of few studies or sparse data, \cite{Takwoingi2017} our findings suggest that model choice was not primarily driven by study numbers. This observation suggests that methodological awareness regarding the appropriate use of statistical models for subgroup analyses could be further improved.
\\
The limited uptake of the HSROC model is not unexpected. Both HSROC and bivariate models can incorporate study-level covariates to explore heterogeneity, but their focus differs: the bivariate model captures heterogeneity in sensitivity and specificity directly, whereas the HSROC framework assesses how study characteristics relate to the position and shape of the entire sROC curve. \cite{Macaskill2023, Trikalinos2012} Because our screening emphasized summary sensitivity and specificity, reviews reporting HSROC results only via AUC or curve parameters may have been under-detected in the present study. 

\subsection{Multiplicity, risk of spurious findings, and prespecification}
Of the meta-analyses with IoH, only \(37/61\) (\(61\%\)) reported testing for differences between subgroups. Those reporting at least one statistically significant subgroup difference examined substantially more subgroup-defining variables overall than those that did not (median \(5\) vs. median \(1.5\); \(p = 0.002\)), whereas the number of primary studies included in the meta-analysis was comparable between groups (median \(11\) vs. median \(12\); \(p = 0.613\)). This suggests that the likelihood of detecting significant subgroup differences may have been driven more by the number of comparisons undertaken than by the strength of the underlying data support. 
Methodological guidance has long emphasized that testing large numbers of subgroups is problematic. IoH in meta-analyses are observational by nature, such that false positive findings become increasingly likely as the number of comparisons increases. To reduce the risk of spurious results, it is recommended to prespecify IoH in study protocols. This approach both limits the number of subgroups investigated and prevents post hoc assessments based on observed study results.\cite{Deeks2024, Dahabreh2016, Deeks2023}
Our findings underline this recommendation. For \(43/61\) (\(70\%\)) of DTA meta-analyses with IoH, a study protocol was available, and within this subset, only \(19/43\) (\(44\%\)) of these studies included fully prespecified IoH. 
(Notably, this proportion is higher than that reported by White et al.\cite{White2024}, who found that only \(13/160\) (\(8.1\%\)) of DTA meta-analyses that conducted subgroup analyses or meta-regression had planned these investigations a priori.)
Investigations that were fully pre-specified assessed markedly fewer subgroup-defining variables (median \(1\)) than those based exclusively on post hoc analyses (median \(5\)) or combining pre-specified and post hoc components (median \(8\); \(p < 0.001\)), while the median number of included primary studies was comparable across these groups (median \(12\), \(12\), and \(13\), respectively; \(p = 0.845\)). 
Although some exploratory post hoc analyses may be justified when supported by external evidence, clear reporting of their status is critical to avoid undue emphasis on potentially spurious findings. \cite{Deeks2024}

\subsection{Strengths and limitations}
This systematic review provides an up-to-date overview of current practice in conducting investigations of heterogeneity in DTA meta-analyses. By sampling recent reviews and applying standardized extraction procedures, we captured a broad cross-section of contemporary methods and reporting practices and interpreted them in light of existing methodological recommendations.  
\\
Nevertheless, some considerations and limitations must be acknowledged.
First, our sample size of \(100\) DTA meta-analyses restricts the ability to detect more nuanced or domain-specific methodological patterns. 
Second, we queried only one database, although the choice of MEDLINE suggests a high-quality source, and our search strategy used specific terminology tailored to DTA meta-analyses, while the eligibility criteria imposed no restrictions regarding target condition or healthcare domain. This resulted in an identified set of records with a high proportion of eligible studies.
Finally, because many reviews included multiple index tests, we selected one test per review to maintain consistency; however, sensitivity analyses suggest that this decision did not materially affect our findings. 
These constraints should be considered when interpreting our results.

\section{Conclusion}
Investigations of heterogeneity remain a common and important component of diagnostic test accuracy meta-analyses, yet their conduct and reporting frequently deviate from established methodological recommendations. In particular, insufficient reporting of statistical models, modest data support, limited prespecification, and extensive exploratory comparisons may compromise the reliability of subgroup findings. Greater methodological awareness, stringent reporting standards, and stronger adherence to prespecified analytical plans are needed to enhance the transparency and robustness of heterogeneity assessment in diagnostic accuracy research.

\backmatter

\section*{List of abbreviations}
\begin{itemize}
    \item IoH: Investigation of heterogeneity
    \item DTA: Diagnostic test accuracy
    \item SGA: Subgroup analysis
    \item MR: Meta-regression
    \item HSROC: Hierarchical summary receiver operating characteristic
    \item OR: Odds ratio
    \item RCT: Randomized controlled trial
    \item IQR: Interquartile range
    \item CI: Confidence interval
    \item AUC: Area under the curve
\end{itemize}

\section*{Declarations}
\subsection*{Ethics approval and consent to participate}
Not applicable.

\subsection*{Consent for publication}
Not applicable.

\subsection*{Availability of data and materials}
The dataset supporting the conclusions of this article is included within the article and its additional file (Additional file 1).

\subsection*{Competing interests}
The authors declare that they have no competing interests.

\subsection*{Funding}
Funded by the Deutsche Forschungsgemeinschaft (DFG, German Research Foundation) – project number 539422711.

\subsection*{Authors' contributions}
LM and AH developed the main research idea and interpreted the study data. LM, AH, and AE performed study selection and data extraction. AF helped develop the systematic search strategy. AH, ZA, AF, and BH contributed to manuscript drafting, provided critical feedback, and reviewed the manuscript for accuracy. All authors read and approved the final manuscript.

\subsection*{Acknowledgements}
We thank Cornelia Fütterer (CF) for assistance with the title and abstract screening.

\section*{Additional files}
\begin{itemize}
    \item File name: Additional file 1
    \item File format: Microsoft Excel-sheet (.xlsx)
    \item Name: Investigation of heterogeneity data
    \item Description: Excel-sheet containing the data underlying this investigation.
\end{itemize}

\begin{appendices}

\section{Systematic Search Strategy}
\label{Systematic Search Strategy}

\textbf{Ovid MEDLINE(R) (December Week 4 2023 to December Week 5 2024)}
Date of latest search: 14 January 2025. 
Results were sorted by Entry Date in descending order using the Ovid interface.

\begin{enumerate}
    \item \texttt{systematic.mp. [mp=ti, ab, sh, hw, tn, ot, dm, mf, dv, kw]}
    \item \texttt{limit 1 to "reviews (best balance of sensitivity and specificity)"}
    \item \texttt{(predict* or diagn* or progn* or screen*).ti,ab.}
    \item \texttt{(test or tests or testing or score).ti,ab.}
    \item \texttt{2 and 3 and 4}
    
    \item \texttt{"diagnostic test accuracy".mp. [mp=ti, ab, sh, hw, tn, ot, dm, mf, dv, kw]}
    \item \texttt{DTA.ti,ab.}
    \item \texttt{accuracy.ti,ab.}
    \item \texttt{meta-analysis.mp. or "meta-analysis as topic"/}
    \item \texttt{6 or 7 or 8 or 9}
    
    \item \texttt{(sensitiv*.tw. and specific*.tw.)}
    \item \texttt{"true negative".tw. and "true positive".tw.}
    \item \texttt{11 or 12}
    
    \item \texttt{5 and 10}
    \item \texttt{5 and 13}
    \item \texttt{14 and 15}
    \item \texttt{limit 16 to (english language and yr="2024")}
\end{enumerate}

\section{Sensitivity analysis and multivariable logistic regression results}
\label{Sensitivity analysis and multivariable logistic regression results}

\begin{table}[h]
\caption{Sensitivity analysis: univariable logistic regression results for diagnostic test accuracy meta-analyses reporting only a single index test with vs.\ without investigations of heterogeneity}
\label{SensAnalysis}
\begin{tabular}{@{}lcccc@{}}
\toprule
Variable & OR & CI & p-value$^{*}$ & N\\
\midrule
Youden Index & 1.20$^{a}$ & (0.95 – 1.56) & 0.135 & 65\\
AUC Value & 1.72$^{a}$ & (0.92 – 3.75) & 0.117 & 41\\
Number of Studies & 1.54$^{b}$ & (1.09 – 2.56) & \textbf{0.044} & 65\\
Number of Patients & 3.91$^{c}$ & (1.35 – 14.28) & \textbf{0.023} & 53\\
AI-related Study & 0.90 & (0.27 – 3.31) & 0.868 & 65\\
Uniform Cut-off & 2.15 & (0.57 – 10.61) & 0.290 & 55\\
Prospective Studies & 1.39 & (0.31 – 5.65) & 0.651 & 46\\
Retrospective Studies & 2.75 & (0.30 – 25.1) & 0.338 & 49\\
Cohort Studies & 4.57 & (0.72 – 39.08) & 0.120 & 29\\
Case-control Studies & 2.14 & (0.49 – 11.54) & 0.331 & 36\\
Cross-sectional Studies & 0.90 & (0.22 – 3.63) & 0.881 & 33\\
\bottomrule
\end{tabular}
\footnotetext{
$^{*}$Wald z-test for univariable logistic regression model. $^{a}$Per 0.1 additional value. $^{b}$Per 5 additional primary studies. $^{c}$Based on a log$_{10}$ transformation of the number of patients; odds ratio reflects a 10-fold increase.
Study design characteristics represent the proportion of index tests for which at least one underlying primary study was of the respective design.
No multiplicity adjustment was considered.
}
\end{table}

\begin{table}[h]
\caption{Multivariable logistic regression results for diagnostic test accuracy meta-analyses with vs.\ without investigations of heterogeneity}
\label{MVAnalysis}
\begin{tabular}{@{}lcccc@{}}
\toprule
Variable & OR & CI & p-value$^{*}$ & N\\
\midrule
Youden index & 1.11$^{a}$ & (0.84 – 1.51) & 0.466 & 54\\
Number of studies & 2.83$^{b}$ & (1.44 – 6.74) & \textbf{0.008} & 54\\
Case-control studies & 3.11 & (0.78 – 14.3) & 0.119 & 54\\
\bottomrule
\end{tabular}
\footnotetext{
$^{*}$Wald z-test for univariable logistic regression. $^{a}$Per 0.1 additional value. 
$^{b}$Per 5 additional primary studies. 
Study design characteristics represent index tests for which at least one underlying primary study was of the respective design.
}
\end{table}

\end{appendices}

\bibliography{sn-bibliography}

\end{document}